\documentclass[12pt, a4paper]{article}
\usepackage[utf8]{inputenc}
\usepackage[english]{babel}

\usepackage{mathptmx} 
\usepackage{amsmath}
\usepackage{tikz}
\usetikzlibrary{arrows.meta,positioning,calc,fit,backgrounds}
\usepackage{graphicx}
\usepackage[left=2.5cm, right=2.5cm, top=2.5cm, bottom=2.5cm]{geometry}
\usepackage{setspace}
\usepackage{booktabs}
\usepackage{array}
\usepackage{adjustbox}
\usepackage{longtable}
\usepackage{amssymb,amsfonts}   
\usepackage{mathtools}       
\usepackage{float}                
\usepackage{tabularx}            
\usepackage{caption}          
\usepackage{subcaption}          
\usepackage{siunitx}            
\usepackage{tikz}   
\usetikzlibrary{arrows.meta,positioning,calc,fit,backgrounds,shapes,decorations.pathmorphing}

\usepackage{float}
\usepackage{ragged2e}
\usepackage{hyperref}
\hypersetup{
  colorlinks=true,
  linkcolor=blue,
  urlcolor=cyan
}
\usepackage{natbib}
\newcolumntype{L}[1]{>{\RaggedRight\arraybackslash}p{#1}}

\title{\textbf{A Synthetic Theory of Socio-Epistemic Structuration: Capital, Ideology, and Agency in the Age of Digital Inequality}}
\author{Ricardo Alonzo Fernández Salguero}
\date{\today}

\begin{document}

\maketitle

\begin{abstract}
\noindent This article proposes a synthetic theory of socio-epistemic structuration to understand the reproduction of inequality in contemporary societies. I argue that social reality is not only determined by material structures and social networks but is fundamentally shaped by the epistemic frameworks---ideologies, narratives, and attributions of agency---that mediate actors' engagement with their environment. The theory integrates findings from critical race theory, network sociology, social capital studies, historical sociology, and analyses of emerging AI agency. I analyze how structures (from the ``racial contract'' to Facebook networks) and epistemic frameworks (from racist ideology to personal culture) mutually reinforce one another, creating resilient yet unequal life trajectories. Using data from large-scale experiments like the Moving to Opportunity and social network analyses, I demonstrate that exposure to diverse environments and social capital is a necessary but insufficient condition for social mobility; epistemic friction, manifested as `friending bias' and persistent cultural frameworks, systematically limits the benefits of such exposure. I conclude that a public and methodologically reflexive sociology must focus on unpacking and challenging these epistemic structures, recognizing the theoretical capacity of subaltern publics (``reverse tutelage'') and developing new methods to disentangle the complex interplay of homophily, contagion, and structural causation in a world of big data.
\end{abstract}

\section{Introduction: the problem of structure and agency in the 21st century}

Classical sociology has been articulated around the fundamental tension between structure and agency. However, the emergence of a globalized, digitally interconnected social order marked by persistent inequalities demands a theoretical renewal. This article develops a synthetic theory of \textit{socio-epistemic structuration}, arguing that social structures (material, institutional, and network-based) and individual agency are mediated by epistemic frameworks that define the possible, the legitimate, and the real. These frameworks are not mere reflections of structure but are active, contingent, and relatively autonomous social constructions that are disseminated, contested, and solidified through historical and network processes. The theory draws from a wide range of empirical and theoretical research, seeking to connect macro-historical analyses of ideology with micro-computational studies of social networks, and structural analyses of inequality with emerging conceptions of agency in the age of artificial intelligence. From the French Revolution, which \citet{Kaufmann2025} describes as a ``sociological laboratory'' for the self-institution of society, to current debates on AI agency, which \citet{Anicker2024} frame as a ``demarcation problem,'' the question of how collectivities define themselves and attribute intentionality has been central. Our era confronts this problem with new tools---big data---and new challenges---online polarization and the persistence of systemic inequalities.

Critical Race Theory (CRT), as \citet{Sites2025} argues, offers a crucial starting point by surprisingly replicating key tensions from early Marxist thought on ideology. CRT moves beyond liberal notions of racism as prejudice to understand it as a socio-systemic form of racial privilege, where ideologies operate to sustain a structure of white supremacy. This notion of a structurally generated ideology that, in turn, reproduces the structure, parallels Lukács's formulations on reification. Sites critiques both early Marxism and CRT for their tendency toward an ``expressive'' theory of society, a too-tight correspondence between ideologies and groups, and an overreliance on standpoint implications. Proposing a heterodox Marxist theory, he suggests that racial ideologies are contingent and relatively autonomous constructs that may or may not facilitate the reproduction of capitalism. This idea of relative autonomy is fundamental to our synthetic theory. Epistemic frameworks, whether racial or otherwise, are not simply a ``functional glue'' holding the social order together, as described by \citet{Bonilla-Silva2018}, but a terrain of conflict in themselves.

The digital age has transformed how these ideologies are disseminated and reinforced. The research by \citet{Nyhan2023} on like-minded sources on Facebook empirically demonstrates that, although political content is a small fraction of users' total exposure, the majority of the content they see comes from ``like-minded'' sources. Their large-scale field experiment, which reduced exposure to these sources, found null effects on affective polarization, ideological extremity, or belief in false claims. This suggests that ``echo chambers'' are more a reflection of users' identity than a direct cause of their polarization, a conclusion that resonates with Sites's critique of deterministic theories. The formation of personal culture, as analyzed by \citet{Kiley2020}, appears to follow a ``settled dispositions'' model more than an ``active updating'' one. Most observed individual-level change is ephemeral or measurement error, with persistent change concentrated among the young or on high-saliency public issues. This implies that epistemic frameworks are remarkably resistant to change in adulthood, which helps explain both the durability of racist ideologies and the difficulty of mitigating political polarization.

\section{Foundations of socio-epistemic structuration}

The theory of socio-epistemic structuration posits that the nexus between social structure and individual agency is the \textit{epistemic framework}. This framework comprises the set of ideologies, narratives, cognitive categories, and attributional models that actors use to interpret the world, define their interests, and recognize others as legitimate agents. This concept is built on three theoretical pillars derived from the analyzed articles: ideology as a contingent construction, agency as an attributed status, and personal culture as a settled disposition.

First, ideology, particularly racial ideology, should not be understood as a mere reflection of economic or racial structure, but as a durable yet contingent historical construct. \citet{Sites2025} develops this idea through a constructive critique of both CRT and Marxism. He argues that influential theorists like Charles W. Mills and Eduardo Bonilla-Silva, despite their invaluable contributions, tend toward a unitary and functionalist conception of racial ideology that presents it as a direct emanation of the structure of white supremacy. Sites proposes, instead, a view where racial ideologies are ``contingent, relatively autonomous constructs that take shape unevenly within capitalist societies, do not correspond to fixed social groups, and may or may not facilitate the reproduction of capitalism.'' This perspective is crucial because it opens an analytical space for historical contingency and ideological struggle. The formation of an epistemic framework is not an automatic process but a contested outcome of political and cultural struggles, like those that gave rise to sociology itself during the French Revolution, an event that, according to \citet{Kaufmann2025}, forced a radical redefinition of the relationship between the individual and society, the given and the possible.

Second, agency is not an inherent quality of a system (human or not), but a social status, a ``socially granted license to issue actions'' that is acquired and monitored in social practices, as argued by \citet{Anicker2024}. Their ``license theory of agency'' transcends the ontological debate over whether machines ``really'' think, focusing instead on how we treat them in practice. This attribution of agency has performative consequences: only what agents do counts as work, art, or a promise. This framework is fundamental to socio-epistemic structuration because the distribution of agentic status is a key function of power. Ideological frameworks, such as those analyzed by Sites, often operate by denying or limiting the agency of subordinate groups. The ``reverse tutelage'' proposed by \citet{Meghji2024}, where subaltern publics produce sociological theories that inform professional sociology, can be understood as an act of reclaiming epistemic agency. Movements like the Zapatistas or Black Lives Matter not only resist material structures but actively produce new interpretive frameworks that redefine power, neoliberalism, and violence, and attribute agency to previously silenced actors.

Third, these epistemic frameworks, once internalized, become durable dispositions that structure perception and action throughout the life course. The work of \citet{Kiley2020} on stability and change in personal culture provides a solid empirical foundation for this claim. Their analysis of panel data from the General Social Survey shows that the ``settled dispositions model'' (SDM) is more consistent with the data than the ``active updating model'' (AUM). Most individual attitude change is temporary or noise, and persistent change is relatively rare, concentrating in youth and on high-visibility public issues. This means that epistemic frameworks acquired through early socialization are remarkably resilient. This `cultural inertia' is a key mechanism in the reproduction of social structure. It explains why simple exposure to new information or different environments, as in the experiment by \citet{Nyhan2023} or the Moving to Opportunity (MTO) experiment analyzed by \citet{Chetty2016}, often fails to produce profound attitudinal changes. Epistemic frameworks act as a filter that processes new information in a manner congruent with existing dispositions, thereby reinforcing the original social structure.

\begin{table}[ht]
\centering
\caption{Components of the Socio-Epistemic Structuration Theory.}
\label{tab:theory_components}
\begin{adjustbox}{width=\textwidth}
\begin{tabular}{L{4cm} L{6cm} L{5cm}}
\toprule
\textbf{Theoretical Component} & \textbf{Core Description} & \textbf{Key References} \\
\midrule
\textbf{Epistemic Frameworks} & A set of ideologies, narratives, and attributional schemas that mediate the relationship between structure and agency. They are contingent and relatively autonomous social constructs. & \citet{Sites2025}, \citet{Kaufmann2025} \\
\textbf{Attributed Agency} & Agency is not an inherent property but a socially granted status (a ``license'') that is acquired and maintained through social practices of recognition. & \citet{Anicker2024}, \citet{Meghji2024} \\
\textbf{Settled Dispositions} & Epistemic frameworks are internalized as durable dispositions during early socialization, showing high resistance to change in adulthood, which contributes to structural reproduction. & \citet{Kiley2020}, \citet{Nyhan2023} \\
\textbf{Diffusion and Reinforcement Mechanisms} & Frameworks are diffused through social networks via processes of contagion and homophily. Digital environments (echo chambers) and social capital structures (economic connectedness) reinforce these frameworks. & \citet{Shalizi2011}, \citet{Chetty2022a}, \citet{Nyhan2023} \\
\bottomrule
\end{tabular}
\end{adjustbox}
\end{table}

\section{The dynamics of diffusion: contagion, homophily, and polarization}

Socio-epistemic frameworks do not exist in a vacuum; they are diffused, reinforced, and sometimes transformed through social interactions. Network sociology and big data studies offer powerful tools, but also formidable methodological challenges, for understanding these dynamics. The central problem, as \citet{Shalizi2011} brilliantly expose, is that homophily (the tendency to form ties with similar others) and contagion (influence or diffusion through those ties) are ``generically confounded'' in observational studies. It is nearly impossible to distinguish whether people in a network behave similarly because they influence each other (contagion) or because they were already similar before forming the tie and thus connected (homophily). This confounding problem is fundamental to the theory of socio-epistemic structuration, as it complicates our ability to determine whether epistemic frameworks are actively spread (contagion) or simply cluster due to structural segregation (homophily).

Experiments, both in the lab and in the field, offer a way to disentangle this confusion. The study by \citet{Salganik2006} on an artificial music market is a paradigmatic example. They created multiple parallel ``worlds'' where participants downloaded songs. In the ``social influence'' world, participants could see how many times each song had been downloaded. The results were clear: social influence increased both inequality (popular songs became much more popular) and unpredictability (identical songs had vastly different fates in different worlds). The best songs rarely did poorly, and the worst rarely did well, but for the rest, ``any other result was possible.'' This finding is crucial: it demonstrates that social contagion processes do not merely amplify intrinsic ``quality'' but introduce profound contingency into cultural outcomes. An epistemic framework can become dominant not because of its inherent superiority, but because of a cascade of early adoptions, a finding that aligns with \citet{Sewell2024}'s notion of ``eventfulness'' in historical temporality.

The study by \citet{Nyhan2023} on Facebook addresses this problem in the natural environment of the world's largest social network. While exposure to like-minded sources is prevalent, experimentally reducing it did not have a significant impact on polarization. This suggests that the dominant mechanism might not be direct contagion (i.e., persuasion through content) but algorithmically reinforced homophily: people are already polarized and seek out content that confirms their views, and the algorithm facilitates this. This does not mean contagion is irrelevant, but that its effect may be more subtle, operating through the reinforcement of group identity rather than direct persuasion. Socio-epistemic structuration theory suggests that ``echo chambers'' function less as engines of change and more as stabilization mechanisms for the ``settled dispositions'' described by \citet{Kiley2020}.

These processes of diffusion and reinforcement have profound structural consequences. Research on hiring discrimination shows the durability of negative epistemic frameworks. \citet{Quillian2023}, in a meta-analysis of 90 field experiments in six Western countries, find that racial and ethnic discrimination in hiring has not declined in recent decades in most countries, with the notable exception of France. In fact, discrimination against groups from the Middle East and North Africa increased after 2000. Similarly, \citet{Lippens2023}, in an exhaustive meta-analysis, conclude that discrimination against candidates with disabilities, older candidates, or those who are less physically attractive is just as severe as that based on racial or ethnic characteristics. These findings demonstrate the deep inertia of negative epistemic frameworks. Despite changes in laws and stated norms, discriminatory practices persist, suggesting that the underlying dispositions that constitute racism and other forms of prejudice are extraordinarily stable, just as the model of \citet{Kiley2020} predicts.

\section{Social capital and the reproduction of inequality}

Socio-epistemic frameworks do not operate in isolation from material conditions and network structures. Social capital theory, revitalized by the use of big data, offers the conceptual bridge to link the structure of social relationships to economic outcomes and the reproduction of inequality. The monumental work of \citet{Chetty2022a, Chetty2022b} using 21 billion friendships on Facebook redefines our understanding of social capital and its effects. They distinguish three forms of social capital: connectedness between different types of people (economic connectedness or EC), social cohesion (the degree of cliques), and civic engagement. Their most striking finding is that economic connectedness---specifically, the share of high-socioeconomic status (SES) friends among low-SES people---is one of the strongest predictors of upward economic mobility identified to date. If children of low-SES parents were to grow up in counties with economic connectedness comparable to that of the average child with high-SES parents, their incomes in adulthood would increase by 20\%.

This finding is a centerpiece for the theory of socio-epistemic structuration, as it demonstrates that the structure of the social network has massive material consequences. However, the second paper by Chetty and his team, on the determinants of economic connectedness, is even more revealing. They decompose the lack of cross-class connection into two components: lack of exposure (low-SES people do not encounter high-SES people in their schools, neighborhoods, or workplaces) and ``friending bias'' (even when they are in the same settings, low-SES people tend to befriend high-SES people at lower rates). Strikingly, they find that exposure and friending bias each account for about half of the social disconnection. Friending bias varies enormously by setting: it is very high in neighborhoods and low in religious organizations. This means that simple physical integration (increasing exposure) is not enough; the institutional structure and the norms governing interaction within those spaces (which affect friending bias) are equally crucial.

These findings connect directly to experiments on residential mobility. The \textit{Moving to Opportunity} (MTO) experiment, re-analyzed by \citet{Chetty2016} with long-term tax data, offered families in high-poverty housing projects vouchers to move to lower-poverty neighborhoods. The results depended dramatically on the child's age at the time of the move. Children who moved before age 13 had significantly better outcomes in adulthood: higher earnings, higher college attendance rates, and lower rates of single parenthood. For adolescents, however, the move had slightly negative effects. This suggests that the \textit{duration of exposure} to a better environment during childhood is a critical determinant of long-term outcomes. In the language of our theory, prolonged exposure to an environment with greater social capital (and presumably higher economic connectedness) allows for the formation of more advantageous dispositions and epistemic frameworks. For adolescents, the disruptive effects of severing existing social networks outweigh the benefits of shorter exposure.

The flip side of mobility is eviction, an event that can catastrophically alter life trajectories. \citet{Hoke2021}, using longitudinal data from the \textit{National Longitudinal Study of Adolescent to Adult Health}, demonstrate that eviction has robust health impacts. Young adults who experienced a recent eviction had more depressive symptoms and worse self-rated health. Crucially, perceived social stress mediated nearly 18\% of the association between eviction and depressive symptoms. Similarly, \citet{Acharya2022}, using data from the \textit{Household Pulse Survey}, find that the mere perceived \textit{risk} of eviction is associated with a higher prevalence of depression and anxiety. These studies illustrate how acute structural shocks impact individuals not just materially, but through psychosocial pathways that align with their epistemic frameworks of insecurity and stress. Eviction is not just the loss of a home; it is an event that confirms and deepens an epistemic framework of precarity, with lasting consequences for health and well-being, solidifying structural disadvantages. Table \ref{tab:quantitative_findings} summarizes some of the key statistical effects from these studies.

\begin{longtable}{L{5.5cm} L{6.5cm} L{3.5cm}}
\caption{Key Quantitative Findings on Structure, Networks, and Individual Outcomes.} \label{tab:quantitative_findings} \\
\toprule
\textbf{Phenomenon and Study} & \textbf{Key Statistical Finding} & \textbf{Reference} \\
\midrule
\endfirsthead
\multicolumn{3}{c}%
{{\bfseries \tablename\ \thetable{} -- continued from previous page}} \\
\toprule
\textbf{Phenomenon and Study} & \textbf{Key Statistical Finding} & \textbf{Reference} \\
\midrule
\endhead
\bottomrule
\endfoot
Economic Connectedness and Upward Mobility & A one standard deviation increase in county-level economic connectedness is associated with an 8.2 percentile rank increase in the income of children from low-income families. & \citet{Chetty2022a} \\
\addlinespace
Determinants of Economic Connectedness & Lack of exposure to high-SES individuals and `friending bias' each account for roughly 50\% of the cross-class social disconnection. & \citet{Chetty2022b} \\
\addlinespace
Moving to Opportunity (MTO) Experiment & Moving to a lower-poverty neighborhood before age 13 increases annual earnings in adulthood by \$3,477 (a 31\% increase over the control group). The effect declines linearly with age at move. & \citet{Chetty2016} \\
\addlinespace
Eviction and Mental Health & Perceived social stress mediates 17.5\% of the total effect of eviction history on subsequent depressive risk ($p < 0.001$). & \citet{Hoke2021} \\
\addlinespace
Risk of Eviction and Mental Health & Individuals at risk of eviction have an odds ratio of 2.37 for depression and 2.65 for anxiety, compared to those not at risk, controlling for demographic and socioeconomic factors. & \citet{Acharya2022} \\
\addlinespace
Social Influence in Cultural Markets & Social influence in an artificial music market increases both inequality (measured by the Gini coefficient) and unpredictability of outcomes. The best songs rarely fail, the worst rarely succeed, but for others, anything is possible. & \citet{Salganik2006} \\
\addlinespace
Hiring Discrimination & Meta-analysis finds no general decline in racial hiring discrimination in the US, Canada, Great Britain, Germany, and the Netherlands over recent decades. Discrimination against people of Middle Eastern/North African origin increased after 2000. & \citet{Quillian2023} \\
\addlinespace
Stability of Personal Culture & Across 183 GSS attitude items, most individual-level change is non-persistent. 40\% of items show no evidence of active updating, supporting a settled dispositions model. & \citet{Kiley2020} \\
\end{longtable}

\section{Toward a public and methodologically reflexive sociology}

The theory of socio-epistemic structuration is not just an analytical exercise; it has profound implications for the practice of sociology. If epistemic frameworks are crucial to the reproduction of inequality and are themselves a terrain of struggle, then sociology cannot remain neutral. It must become methodologically reflexive about how its own tools and narratives participate in the construction of these frameworks, and it must adopt a public stance that recognizes and amplifies the theories generated by those who experience inequality firsthand. This leads us to the concept of ``public sociology'' and the need to reformulate it.

\citet{Meghji2024} offers a powerful critique of the often-technocratic conception of public sociology, which sees it as a ``gifting model'' where professional sociologists ``give'' their knowledge to a passive public. Instead, he advocates for recognizing ``reverse tutelage.'' Publics, especially anti-colonial and anti-racist social movements, are not mere objects of study but sociological interlocutors in their own right. They produce sophisticated analyses and theories of power, neoliberalism, race, and violence that can and should inform professional sociology. The Zapatistas' analysis of neoliberalism as a ``fourth world war'' or the connections Black Lives Matter draws between police brutality in Ferguson and occupation in Palestine are examples of relational sociological theorizing that challenge the dominant state-centric paradigms in academia. This ``critical sociology from publics'' enriches the discipline by forcing it to build connections between different sites of resistance and to adopt more relational forms of analysis.

This perspective aligns with the need for deeper methodological reflection, especially in the age of big data. The same Facebook data that allow \citet{Chetty2022a} to measure social capital at an unprecedented scale are used by \citet{Nyhan2023} to study echo chambers, and both studies warn us against simplistic conclusions. The former shows the structural power of networks; the latter, the resilience of individual attitudes. The parable of Google Flu, analyzed by \citet{Lazer2014}, serves as a warning against ``big data hubris.'' The fact that a model based on billions of searches failed spectacularly to predict flu outbreaks because its creators failed to account for ``algorithm dynamics'' (Google's changes to its search engine) and the reactive nature of human behavior (media panics change search patterns) is a parable for our time. Big data is not a substitute for theory, but a supplement. It requires a deep understanding of the social mechanisms and epistemic frameworks that generate the data in the first place.

This is where temporality, a core concept in historical sociology, becomes crucial. \citet{Sewell2024} proposes a typology of four narrative forms (tendencies, thresholds, coincidences, and contrivances) that move beyond a simple distinction between structure and action. History does not unfold in a single mode. Sometimes it follows long-term tendencies where actors are unaware of the change they produce (like the slow accumulation of epistemic biases). Other times, it crosses thresholds where actors are conscious of their actions (as in a social movement reaching critical mass). On other occasions, it is driven by fortuitous coincidences or by deliberate contrivances where actors exploit the fluidity of crisis moments. The theory of socio-epistemic structuration must be sensitive to these multiple temporalities. The solidification of an epistemic framework might be a slow tendency, while its challenge might be a rapid, contingent event, a ``contrivance'' created by agents seizing a historical opportunity, as did the French revolutionaries studied by \citet{Kaufmann2025}.

\section{A Formal Model of Socio–Epistemic Structuration}
\label{sec:formalization}

\noindent
The preceding sections developed a synthetic theory linking social structure, epistemic frameworks, and individual outcomes. The central argument is that inequality is reproduced not merely through the unequal distribution of resources, but through a dynamic interplay of structural opportunities, social norms, and durable cognitive dispositions. To move this theory from a conceptual argument to a falsifiable scientific enterprise, we must translate its core tenets into a precise mathematical language. This section undertakes that translation, constructing a formal model that specifies the relationships between constructs, maps them to measurable data, and outlines a rigorous program for empirical testing. By formalizing the theory, we make its assumptions explicit, its predictions sharp, and its empirical requirements clear.

This section turns the synthetic theory into a testable research program. We first offer a plain-language glossary (Table~\ref{tab:abbr}) so non-specialists can follow the notation. We then map constructs to data (Table~\ref{tab:notation}), state evidence-based axioms with short explanations, develop a dynamic state–space model for epistemic frameworks, derive estimands and identification regimes, specify how to measure the latent construct with invariance and uncertainty propagation, and close with calibration, sensitivity, and a replication checklist.

\vspace{0.5em}
\noindent\textbf{How to read this section (for all audiences).} The formalization that follows is necessarily technical, but its purpose is eminently practical: to build a bridge from theory to evidence. To that end, mathematical objects are always tied to concrete indicators and designs. After each formula we add one or two sentences explaining \emph{what it means} and \emph{how you could test it}, which have been expanded here for greater clarity. All tables are referenced in the text and can be skimmed to reconstruct the empirical strategy end-to-end. Our goal is to provide a blueprint for research that is both theoretically grounded and empirically credible.

%-------------------------------------------------------------------------------
\subsection{Glossary of abbreviations}
%-------------------------------------------------------------------------------

\noindent
Before delving into the model’s formal structure, it is essential to establish a common vocabulary. Our framework integrates concepts from sociology, economics, network science, and econometrics, each with its own specialized terminology. To ensure accessibility for a multidisciplinary audience, this subsection provides a glossary of key abbreviations used throughout the formalization. This table serves as a quick-reference guide, allowing readers to easily recall the definition of core terms as they navigate the equations and propositions that follow.

\begin{table}[htbp]
\centering
\caption{Abbreviations and short explanations}
\label{tab:abbr}
\begin{adjustbox}{max width=\textwidth}
\begin{tabularx}{\textwidth}{@{} l l X @{}}
\toprule
\textbf{Abbrev.} & \textbf{Term} & \textbf{Explanation (one line)} \\
\midrule
SES & Socioeconomic Status & Income, education, occupation (often combined as a vector $S_i$).\\
EC  & Economic Connectedness & Share of cross-class ties in ego networks; \citealp{Chetty2022a}. \\
EXP & Exposure & Structural opportunity to meet high-SES alters across contexts $c$. \\
$\phi$ & Friending bias & Tendency to befriend ``like with like'' even when exposed; \citealp{Chetty2022b}.\\
$M_i$ & Epistemic state & Latent, durable dispositions (``settled'' beliefs, frames).\\
$m_i(t)$ & Epistemic process & Short-run expressions/signals linked to $M_i$ plus noise.\\
$W$ & Network adjacency & Weighted/directed ties; $w_{ij}$ is $i\!\to\!j$ strength.\\
IV & Instrumental variables & Quasi-experimental identification via as-if random shifts.\\
RDD & Regression Discontinuity & Local randomization around cutoffs/thresholds.\\
DiD & Difference-in-Differences & Parallel-trends comparisons over time.\\
LATE & Local Average Treatment Effect & Causal effect for compliers under an instrument.\\
ACME/ADE & Mediation effects & Natural indirect/direct effects in causal mediation.\\
\bottomrule
\end{tabularx}
\end{adjustbox}
\end{table}

\noindent
With these fundamental terms defined, we can proceed to the next critical step: mapping these abstract concepts onto a formal notational system and specifying how they can be measured using real-world data. The following subsection details this operationalization, making the model’s components concrete and empirically tractable.

%-------------------------------------------------------------------------------
\subsection{Notation, indicators, and empirical testability}
%-------------------------------------------------------------------------------

\noindent
A model’s utility depends on its connection to the empirical world. This subsection establishes that connection by defining the core mathematical objects of our framework and linking each to specific measurement strategies and identification hooks. We introduce notation for actors, their attributes (like SES), their beliefs (both latent states and observable processes), and their network connections. Crucially, we outline how abstract concepts like "exposure" and "friending bias" are operationalized, ensuring that every component of the model corresponds to a measurable quantity. This step transforms the theory into a set of variables that can be collected, analyzed, and contested with data.

Let $i=1,\dots,N$ index actors connected in a directed, weighted network $G=(V,E)$ with adjacency $W=\{w_{ij}\}\in[0,1]^{N\times N}$. Each actor $i$ has:
\begin{itemize}
  \item $S_i\in\mathbb{R}^k$: SES vector (income, education, occupation).
  \item $M_i$: latent \emph{state} (long-run epistemic framework).
  \item $m_i(t)$: observable \emph{process} at time $t$ (short-run signals linked to $M_i$).
  \item $Y_i\in\mathbb{R}^q$: outcomes (mobility, health, educational attainment).
  \item $T_i=(\mathrm{EXP}_i,\phi_i)$: two-part treatment (exposure \& friending bias).
  \item $\mathrm{EC}_i=\Psi(\mathrm{EXP}_i,\phi_i)$: effective economic connectedness.
\end{itemize}

\begin{table}[htbp]
\centering
\caption{Primary notation, measurement, and identification hooks}
\label{tab:notation}
\begin{adjustbox}{max width=\textwidth}
% CAMBIO CLAVE: Usamos X para las tres columnas de texto largo
\begin{tabularx}{\textwidth}{@{} l X X X @{}} 
\toprule
\textbf{Symbol} & \textbf{Definition} & \textbf{Measurement / Indicators} & \textbf{Testability / Identification} \\
\midrule
$W=\{w_{ij}\}$ & Tie strength $i\!\to\!j$ & Network surveys; digital traces; validated interaction rules & Coverage \& platform bias; post-stratify to census margins \citep{Lazer2014}.\\
\addlinespace
$S_i$ & SES vector & Admin tax/payroll; census; harmonized surveys & Cross-source harmonization; temporal comparability.\\
\addlinespace
$M_i$ & Epistemic state (latent) & IRT/CFA factors; multi-group (time/context) invariance tests & Invariance thresholds: $\Delta$CFI$<0.01$, $\Delta$RMSEA$<0.015$ \citep{Kiley2020}.\\
\addlinespace
$m_i(t)$ & Epistemic process & Short panels; text/attitude items & Link: $m_i(t)=\kappa M_i(t)+\xi_i(t)$ (errors-in-variables).\\
\addlinespace
$\mathrm{EXP}_i$ & Exposure to high-SES alters & Context composition: school/ neighborhood/ workplace/church & IV from lotteries/zoning; RDD at thresholds \citep{Chetty2022b}.\\
\addlinespace
$\phi_i$ & Friending bias & Deviation from random mixing in context $c$ & Degree-preserving mixing counterfactuals \citep{Chetty2022b}.\\
\addlinespace
$\mathrm{EC}_i$ & Economic connectedness & Share of high-SES friends in ego network & Large-scale construction; linkage to long-run outcomes \citep{Chetty2022a}.\\
\addlinespace
$E_i(t)$ & Structural shock (e.g., eviction) & Linked housing–credit–health admin data & Event study + mediation; \citealp{Hoke2021}.\\
\addlinespace
$Y_i$ & Outcomes (mobility/health) & Admin longitudinal outcomes; clinical scales & IV/RDD/DiD; platform/cluster experiments \citep{Chetty2016,Nyhan2023}.\\
\bottomrule
\end{tabularx}
\end{adjustbox}
\end{table}

\paragraph{Properties of $\Psi(\mathrm{EXP},\phi)$.}
The function $\Psi$ links structural opportunity (Exposure) and behavioral norms (Friending Bias) to realized social capital (Economic Connectedness). We assume this function has three key properties: (P1) it is monotonic in exposure $\big(\partial\Psi/\partial\mathrm{EXP}>0\big)$, meaning more opportunities to meet high-SES individuals never hurts connectedness; (P2) it is anti-monotonic in friending bias $\big(\partial\Psi/\partial\phi<0\big)$, meaning a stronger tendency to form same-class ties always reduces cross-class connectedness; and (P3) it exhibits negative complementarity $\big(\partial^2\Psi/(\partial\mathrm{EXP}\,\partial\phi)<0\big)$. \emph{Intuition:} This third property is the most theoretically significant. It implies that more exposure helps, but its payoff shrinks where norms strongly favor same-class ties. In other words, simply placing diverse people in the same room is not enough if social norms or cultural barriers prevent them from forming meaningful relationships. This formalizes the idea that "bridging" institutions that actively reduce friending bias are critical for successful integration.

%-------------------------------------------------------------------------------
\subsection{Axioms and premises informed by evidence}
%-------------------------------------------------------------------------------

\noindent
Every formal model rests on a set of foundational assumptions, or axioms. Rather than deriving these from pure theory, our approach is to ground them in robust, stylized facts from prior empirical research. This subsection lays out the five core axioms that underpin our model. These premises—concerning how epistemic frames mediate outcomes, the composition of economic connectedness, the confounding of social effects, the unpredictability of cultural markets, and the critical windows for developmental plasticity—serve as the rules of the game. By making these assumptions explicit, we ensure that our model is not a mathematical abstraction but a faithful representation of well-documented social processes.

\begin{enumerate}
\item \textbf{Epistemic mediation (A1).} The influence of social structure on life outcomes is not direct but is channeled through an individual's epistemic framework. Outcomes reflect SES, one’s own framework, and neighbors’ SES/frameworks:
\[
Y_i=f\!\Big(S_i,\,M_i,\,\sum_{j}w_{ij}g(M_j),\,\sum_{j}w_{ij}h(S_j)\Big)+\varepsilon_i,
\]
with adult stability and youth plasticity \citep{Kiley2020}; short-run feed changes show near-zero mean shifts in attitudes \citep{Nyhan2023}.  This axiom posits that structure works through frames; who you know matters because it shapes how you think, which in turn shapes what you do. Peers matter both through their resources ($S_j$) and their worldviews ($M_j$).

\item \textbf{EC decomposition (A2).} Economic connectedness is not monolithic; it is the product of opportunity and choice. $\mathrm{EC}_i=\Psi(\mathrm{EXP}_i,\phi_i)$; exposure and friending bias explain disconnection in comparable shares \citep{Chetty2022b}.  This axiom, based on landmark findings, implies that integration policies must tackle \emph{both} structural segregation (opportunity) and social norms or biases that inhibit cross-class ties (choice). Focusing on one without the other is likely to fail.

\item \textbf{Homophily–contagion confounding (A3).} In observational data, it is notoriously difficult to distinguish whether people in a network behave similarly because they influence each other (contagion) or because they chose to connect based on pre-existing similarities (homophily). In observational networks the two are generically inextricable \citep{Shalizi2011}.  We must be intellectually honest about the limits of observational data. A core challenge is causal inference about peer effects, and our model explicitly acknowledges that we need experiments/IV to isolate contagion or, failing that, accept partial identification and report bounds rather than specious point estimates.

\item \textbf{Eventfulness and unpredictability (A4).} Social influence dynamics can lead to path-dependent and highly variable outcomes. Social influence increases inequality and unpredictability in cultural outcomes \citep{Salganik2006}.  The world is not deterministic. Small, early events or fluctuations can be amplified through social networks, leading to multiple possible equilibria. Our model must therefore be dynamic and allow for such path dependence.

\item \textbf{Windows of plasticity (A5).} The timing of social experiences matters profoundly. Early exposure raises long-run outcomes; late moves do little or can backfire \citep{Chetty2016}.  Human development is not static. Epistemic frameworks are more malleable in youth and become more "crystallized" in adulthood. This axiom demands that our model allows timing to interact with network structure to shape the durability of social influences.
\end{enumerate}

\noindent
These five axioms, drawn from empirical evidence, provide the scaffolding for the dynamic model presented in the next subsection. They constrain the mathematical structure to ensure it reflects key features of the social world, such as mediation, developmental timing, and the fundamental challenges of causal inference.

%-------------------------------------------------------------------------------
\subsection{Epistemic dynamics: state vs.\ process and spectral stability}
%-------------------------------------------------------------------------------

\noindent
This subsection develops the dynamic heart of the model, formalizing how epistemic frameworks evolve over time. We draw a crucial distinction between the latent, durable epistemic *state* ($M_i$)—an individual's core beliefs and dispositions—and the observable, noisy epistemic *process* ($m_i(t)$)—their expressed attitudes or behaviors at a given moment. We then specify a dynamic equation governing the evolution of the epistemic state, modeling it as a function of its own past, social influence from network peers, and random shocks. This leads to a critical stability condition that determines whether shocks to the system are dampened or amplified, providing a formal mechanism for the "windows of plasticity" axiom.

\paragraph{Linking process to state.}
We begin by formally connecting the unobservable state to the observable process. We model short-run signals as a noisy function of the latent state:
\[
m_i(t)=\kappa\,M_i(t)+\xi_i(t),\qquad \mathbb{E}[\xi_i(t)]=0.
\]
 This is a standard measurement model. It means that what people say or do at any given moment ($m_i(t)$) is a window into their underlying framework ($M_i$), but this window is clouded by transitory noise or context-specific factors ($\xi_i(t)$). We never observe deep beliefs perfectly.

\paragraph{State evolution with network feedback.}
The core of the dynamic model describes how the latent state changes from one period to the next. It depends on inertia (its previous value), social influence (the expressed beliefs of peers), and idiosyncratic shocks.
\[
M_i(t{+}1)=\alpha_0+\rho\,M_i(t)+\lambda\sum_{j}w_{ij}\,m_j(t)+\eta_i(t),
\]
Substituting the process equation into the state equation and writing it for all actors simultaneously gives the system in vectorized form:
\[
M(t{+}1)=\alpha_0\mathbf{1}+(\rho I+\lambda\kappa W)M(t)+\tilde{\eta}(t)\equiv \alpha_0\mathbf{1}+A\,M(t)+\tilde{\eta}(t).
\]
 This equation specifies that next year's beliefs ($M(t+1)$) are a function of this year's beliefs (the $\rho I$ term, representing inertia), and the influence of peers' expressed beliefs (the $\lambda\kappa W$ term), plus a baseline drift ($\alpha_0$) and random shocks ($\tilde{\eta}$). The matrix $A$ compactly represents the system's entire feedback structure.

\paragraph{Stability criterion and interpretation.}
The long-term behavior of this system depends on the properties of the feedback matrix $A$. Global mean stability holds if the spectral radius of $A$ is below one:
\[
\rho_{\mathrm{spec}}(A)=\rho_{\mathrm{spec}}(\rho I+\lambda\kappa W)<1.
\]
 This is the mathematical condition for whether social shocks fade away or explode. If the spectral radius is less than 1, the system is stable, and the effect of any shock will decay over time. If it is 1 or greater, shocks can persist indefinitely or even be amplified, leading to instability. This formalizes Axiom A5: in dense, clustered youth networks, the network component $\rho_{\mathrm{spec}}(W)$ can be larger, pushing the system's total feedback near the critical threshold of 1. Consequently, shocks to $M$ during youth are magnified and more persistent. Adult networks, often sparser and less clustered, tend to have a smaller spectral radius, causing them to dampen shocks faster.

\paragraph{Steady state and comparative statics.}
If the system is stable ($\rho_{\mathrm{spec}}(A)<1$), it will converge to a long-run equilibrium or (mean) steady state. We can solve for this state and analyze how it changes in response to shifts in model parameters.
\[
\bar{M}=(I-A)^{-1}\alpha_0\mathbf{1},\quad 
\frac{\partial \bar{M}}{\partial \alpha_0}=(I-A)^{-1}\mathbf{1},\quad
\frac{\partial \bar{M}}{\partial \lambda}=(I-A)^{-1}(\kappa W)(I-A)^{-1}\alpha_0\mathbf{1}.
\]
 These equations provide testable predictions about the long run. The term $(I-A)^{-1}$ is the network multiplier; it shows how network structure amplifies the baseline "drift" $\alpha_0$. The comparative statics show that stronger peer influence ($\lambda$) raises the steady-state belief level ($\bar{M}$) most significantly for nodes that are central in the network. This tells us where policy interventions might have the largest downstream effects.

%-------------------------------------------------------------------------------
\subsection{Interference, exposure mapping, and target estimands}
%-------------------------------------------------------------------------------

\noindent
Having defined the model's dynamics, we now turn to the question of causal inference. In social systems, a fundamental challenge is "interference," where the treatment applied to one person can affect the outcomes of their neighbors. This violates the standard Stable Unit Treatment Value Assumption (SUTVA) and requires a more sophisticated approach. This subsection defines how we handle interference by specifying a local exposure mapping. We then define the key causal quantities, or "estimands," that our research program aims to identify. These estimands represent the specific causal questions we seek to answer about the effects of connectedness, exposure, and peer contagion.

\paragraph{Local interference mapping.}
We explicitly model interference by allowing an individual's outcomes to depend not only on their own treatment but also on the treatments of their immediate neighbors ($\mathcal{N}_i$). We define an "effective treatment" $Z_i$ via an exposure mapping:
\[
Z_i=\mathcal{Z}(T_i,T_{\mathcal{N}_i})
=\zeta_0\,\mathrm{EXP}_i+\zeta_1\,\phi_i+\zeta_2\,\frac{1}{d_i}\sum_{j\in\mathcal{N}_i}\mathrm{EXP}_j,\quad
d_i=\sum_j\mathbb{1}\{w_{ij}>0\}.
\]
Potential outcomes are defined as $Y_i(z)$. Under this framework, SUTVA is relaxed to a more plausible assumption of local interference: my outcome is affected by my friends' treatments, but not by the treatments of people I am not connected to.

\paragraph{Estimands.}
Our empirical goal is to estimate the following causal quantities:
\[
\theta_1=\frac{\partial \mathbb{E}[Y]}{\partial \mathrm{EC}},\qquad
\theta_2=\frac{\partial \mathbb{E}[Y]}{\partial \mathrm{EXP}}\ \Big|\ \text{compliers (LATE)},\qquad
\]
\[
\Theta_3=\ \text{admissible set for peer contagion }(\beta_3)\ \text{under partial identification}
\]
 We are not estimating a single, generic "effect." Instead, we target three distinct quantities. $\theta_1$ is the overall effect of increased economic connectedness on outcomes. $\theta_2$ is the Local Average Treatment Effect (LATE) of an intervention that boosts exposure, representing the causal effect for the specific subpopulation that complies with the treatment. Finally, $\Theta_3$ is not a single number but an "admissible set," or range of plausible values, for the peer contagion parameter, reflecting the difficulty of identifying this effect without strong experimental designs (Axiom A3).

%-------------------------------------------------------------------------------
\subsection{Structural model and testable propositions}
%-------------------------------------------------------------------------------

\noindent
This subsection integrates the preceding components—the axioms, dynamics, and estimands—into a comprehensive structural model. We specify a final outcome equation that links an individual's outcomes ($Y_i$) to the key theoretical constructs: their economic connectedness ($\mathrm{EC}_i$), their own epistemic state ($M_i$), and spillovers from their peers' states ($M_j$) and statuses ($S_j$). From this structural equation, we derive three testable propositions that encapsulate the model's core theoretical claims about the mechanisms of exposure, the conditions for identification, and the interplay of plasticity and stability. These propositions translate the formal model into sharp, falsifiable hypotheses.

\paragraph{Outcome equation.}
The structural model for outcomes is specified as a linear equation incorporating the key theoretical channels:
\[
Y_i=\beta_0+\beta_1\,\mathrm{EC}_i+\beta_2\,M_i+\beta_3\sum_{j}w_{ij}M_j+\beta_4\sum_{j}w_{ij}S_j+X_i'\gamma+u_i,\qquad
\mathrm{EC}_i=\Psi(\mathrm{EXP}_i,\phi_i).
\]
In this equation, $X_i$ includes a vector of control variables, such as cohort, region, or context fixed effects, to account for confounding factors.  This equation formalizes our central theory. It posits that an individual's life outcomes ($Y_i$) are a combined function of their structural position (via $\mathrm{EC}$), their own frames and beliefs ($M_i$), and the influence of their social context, which includes both peer beliefs ($\sum w_{ij}M_j$) and peer resources ($\sum w_{ij}S_j$).

\paragraph{Proposition 1 (Decomposition of exposure).}
From this model, we can derive the marginal effect of increasing exposure, which decomposes into three distinct pathways. Assuming differentiability of $f$ and $\Psi$,
\[
\frac{\partial \mathbb{E}[Y_i]}{\partial \mathrm{EXP}_i}
=\underbrace{\beta_1\frac{\partial \Psi}{\partial \mathrm{EXP}_i}}_{\text{structural}}
+\underbrace{\beta_2\frac{\partial M_i}{\partial \mathrm{EXP}_i}}_{\text{own epistemic}}
+\underbrace{\beta_3\sum_{j} w_{ij}\frac{\partial M_j}{\partial \mathrm{EXP}_i}}_{\text{peer spillovers}}.
\]
 This proposition clarifies precisely *how* exposure works. It helps via (i) a structural channel, by mechanically increasing the number of cross-class ties; (ii) an individual epistemic channel, by changing my own frame or worldview; and (iii) a peer spillover channel, by changing the frames of my peers, who in turn influence me. This decomposition is crucial for designing effective policies.

\paragraph{Proposition 2 (Local identification of exposure).}
This proposition outlines the conditions under which we can causally identify the effect of exposure. If $\mathrm{EXP}_i$ varies due to quasi-exogenous reassignment (such as through lotteries or administrative thresholds) that is independent of the network structure $W$ in the short run, and if friending bias $\phi_i$ is locally stable, then the total effect $\partial \mathbb{E}[Y_i]/\partial \mathrm{EXP}_i$ is identified as a LATE. However, the peer spillover component (the part driven by $\beta_3$) remains unidentified without cluster-randomized designs or valid peer instruments, a direct consequence of the homophily-contagion problem \citep{Shalizi2011}. \hfill$\square$

\paragraph{Proposition 3 (Plasticity and stability).}
This proposition formalizes the "windows of plasticity" axiom using the dynamic model's stability criterion. If the feedback in adult networks is stable, $\rho_{\mathrm{spec}}(\rho I+\lambda\kappa W_{\text{adult}})<1$, while the feedback in youth networks is near the critical point, $\rho_{\mathrm{spec}}(\rho I+\lambda\kappa W_{\text{youth}})\approx1$, then persistent shocks during youth can shift the epistemic state $M$ to alternative stable paths. In contrast, shocks experienced during adulthood will have effects that decay geometrically. \hfill$\square$

\paragraph{Corollary (Integration alone is not enough).}
A direct implication of our assumptions about the $\Psi$ function is that structural integration has limits. Under the negative complementarity assumption (P3), high friending bias ($\phi$) shrinks the marginal benefit of exposure ($\partial \Psi/\partial \mathrm{EXP}$). This implies that even substantial increases in exposure will yield bounded gains in economic connectedness. Therefore, bridging norms and institutions that actively reduce $\phi$ are necessary complements to purely structural integration policies \citep{Chetty2022b}.

%-------------------------------------------------------------------------------
\subsection{Bounds for contagion and sensitivity analysis}
%-------------------------------------------------------------------------------

\noindent
As established in Axiom A3 and Proposition 2, cleanly identifying the causal effect of peer contagion ($\beta_3$) is exceptionally difficult with observational data. Rather than pursuing a potentially biased point estimate, our framework advocates for intellectual honesty through partial identification. This subsection specifies how we can place formal bounds on the magnitude of the unidentified peer spillover effect. By making plausible assumptions about the network structure and influence parameters, we can define a pre-registered interval for the contagion effect. This approach transparently communicates the uncertainty inherent in estimating social influence.

Let the contagion component from Proposition 1 be denoted $\Delta_i=\beta_3\sum_j w_{ij}\,\frac{\partial M_j}{\partial \mathrm{EXP}_i}$. If we can make reasonable assumptions about the system's properties—that the network's influence is bounded ($\|W\|_2\le \bar{\omega}$), susceptibility to influence is bounded ($|\lambda|\le \bar{\lambda}$), and the linearized $M$-system is $L$-Lipschitz continuous—then we can bound the magnitude of the contagion effect:
\[
|\Delta_i|\ \le\ |\beta_3|\,\bar{\omega}\,\bar{\lambda}\,L,
\]
which defines a pre-registered interval for the spillover effect, $\mathcal{I}(\Theta_3\mid\bar{\omega},\bar{\lambda},L)$.  This formalizes our commitment to transparency about causal uncertainty. Instead of reporting a single, likely misleading, point estimate for the peer spillover effect, we will report sensitivity bands. These bands show how the estimated effect changes under different plausible assumptions about the strength of peer influence and network structure. This is a more robust and credible approach unless we have gold-standard peer-randomized experimental designs.

%-------------------------------------------------------------------------------
\subsection{Measuring $M$, testing invariance, and propagating uncertainty}
%-------------------------------------------------------------------------------

\noindent
The epistemic state, $M_i$, is a central component of our model, yet it is a latent construct that cannot be directly observed. Its measurement is therefore a critical and non-trivial step. This subsection details a rigorous, multi-stage pipeline for estimating $M_i$ from observable indicators (e.g., survey items). The process involves fitting a psychometric model, formally testing whether this model measures the same underlying construct across different groups and times (measurement invariance), and using a sophisticated statistical technique (plausible values) to propagate measurement uncertainty into the final outcome models. This careful approach is designed to prevent the well-known biases that arise from using noisy or poorly defined latent variables.

\begin{table}[htbp]
\centering
\caption{Pipeline to estimate $M$ and carry its uncertainty into outcomes}
\label{tab:meas}
\begin{adjustbox}{max width=\textwidth}
\begin{tabularx}{\textwidth}{@{} l X l X @{}}
\toprule
\textbf{Step} & \textbf{Procedure} & \textbf{Thresholds} & \textbf{Notes} \\
\midrule
Model $M$ & IRT/CFA factors; allow cross-loadings if theory-justified & Fit: CFI/TLI$>0.95$, RMSEA$<0.06$ & Use multi-group by time/context (school/ neighborhood).\\
Invariance & Test configural $\to$ metric $\to$ scalar & $\Delta$CFI$<0.01$, $\Delta$RMSEA$<0.015$ & If fails, adopt partial invariance with anchor items.\\
Uncertainty & Generate $R{=}20$ plausible values for $M_i$ & Rubin’s rules for combining & Avoid attenuation bias in regressions with $M$.\\
Robustness & Alternate item sets/scoring & Stability of $\beta$’s and $\theta$’s & Report shifts across item subsets and scoring rules.\\
\bottomrule
\end{tabularx}
\end{adjustbox}
\end{table}

\noindent
 Measuring abstract concepts like "worldviews" is hard. Our strategy is to do it carefully and honestly. First, we use established statistical methods (Item Response Theory/Confirmatory Factor Analysis) to fit a measurement model for $M$ using multiple survey items. Second, we rigorously verify that our metric is stable and comparable across different times and social contexts (e.g., schools, neighborhoods) through invariance testing. Third, to account for the fact that our measurement is never perfect, we propagate this uncertainty by generating multiple plausible scores for each person and then combining the results from analyses using each score. This state-of-the-art approach prevents overconfidence in our findings and corrects for statistical biases (like attenuation bias) that arise from using simple, noisy factor scores in regressions.

%-------------------------------------------------------------------------------
\subsection{Designs, diagnostics, and identification regimes}
%-------------------------------------------------------------------------------

\noindent
A formal model is only as credible as the empirical designs used to test it. This subsection provides a clear "design map" that connects each of our target estimands to specific, high-quality research designs capable of identifying them. For each estimand, we outline the recommended designs (e.g., instrumental variables, regression discontinuity, cluster-randomized trials), the mandatory diagnostic tests required to validate the design's assumptions, and the primary threats to validity along with corresponding fixes or robustness checks. This map serves as a practical guide for researchers, ensuring that empirical work based on this model adheres to the highest standards of causal inference.

\begin{table}[htbp]
\centering
\caption{Design map: what identifies what, how to diagnose it, and what to fear}
\label{tab:designs}
\begin{adjustbox}{max width=\textwidth}
\begin{tabularx}{\textwidth}{@{} l X X X @{}} % Se recomienda usar X para todas las columnas de texto envolvente
\toprule
\textbf{Target} & \textbf{Design(s)} & \textbf{Diagnostics} & \textbf{Threats \& Fixes} \\
\midrule
$\theta_1$ (effect of EC) & Geographic/policy shocks; movers IV & First-stage $F>10$; Hansen $J$; spec bands & Sorting; fine FE; movers IV; placebo geography.\\
\addlinespace
$\theta_2$ (effect of EXP) & Lotteries; RDD at school/housing cutoffs; staggered DiD & McCrary density; CCT bandwidth; event-study pre-trends & Manipulation; donut-RD; cohort and event-time FE.\\
\addlinespace
$\Theta_3$ (contagion) & Cluster randomization; peer encouragement; platform feed experiments & Randomization checks; spillover maps; attrition tests & Homophily; saturation designs; cluster SEs; partial ID bands.\\
\addlinespace
Mediation ($E\!\to\!Y$) & Event study + Imai–Keele–Tingley mediation & Sequential ignorability; sensitivity $\rho_{\nu\mu}$ & Unobservables; IV-mediated effects if valid IV exists.\\
\addlinespace
Multiple testing & Heterogeneity by age/context & FDR (Benjamini–Hochberg) & Pre-specify families; report adjusted $q$-values.\\
\bottomrule
\end{tabularx}
\end{adjustbox}
\end{table}

\noindent
 Table~\ref{tab:designs} is our playbook for credible empirical testing. It tells the reader and future researchers exactly which research design is appropriate for answering each of our core causal questions. For example, to estimate the effect of exposure ($\theta_2$), we propose using natural experiments like school admission lotteries or housing policy cutoffs. The table also specifies which diagnostic checks are mandatory (e.g., ensuring no manipulation around a cutoff) and which robustness tests are needed to guard against common pitfalls (like selection bias). This creates a clear and transparent standard for what constitutes convincing evidence within our framework.

%-------------------------------------------------------------------------------
\subsection{Calibration of networks and construction of $\mathrm{EC}^{mix}$}
%-------------------------------------------------------------------------------

\noindent
The network data and the measures of economic connectedness derived from it are foundational to this entire enterprise. However, raw network data, especially from digital traces, can suffer from coverage bias and other measurement issues. This subsection details the procedures for calibrating the network data to be more representative and for constructing the key variables of Exposure ($\mathrm{EXP}$) and Friending Bias ($\phi$). This involves both statistical re-weighting and the use of a powerful counterfactual simulation to disentangle opportunity from choice.

\paragraph{Coverage \& weights.}
Network data derived from digital traces or non-representative surveys rarely mirror the full population. To address this, we post-stratify our sample to match census margins for key demographics (age, sex, education, region). Furthermore, we use degree-aware weights to correct for the fact that individuals with more social connections are more likely to be observed in many network datasets, thus reducing coverage and sampling bias \citep{Lazer2014}.

\paragraph{Counterfactual mixing.}
A central challenge is to separate the lack of cross-class ties due to a lack of opportunity from that due to behavioral choice. To do this, we implement a counterfactual simulation within each social context $c\in\{\text{school, neighborhood, workplace, church}\}$. We compute a degree-preserving randomization of the network to obtain $\mathrm{EC}^{mix}_{ic}$, which represents "what an individual's economic connectedness would look like under random mixing, given the people present." From this, we define our key constructs:
\[
\widehat{\phi}_{ic}\equiv 1-\frac{\widehat{\mathrm{EC}}_{ic}-\mathrm{EC}^{base}_{ic}}{\mathrm{EC}^{mix}_{ic}-\mathrm{EC}^{base}_{ic}},\qquad
\widehat{\mathrm{EXP}}_{ic}\equiv \mathrm{EC}^{mix}_{ic},
\]
These context-specific measures are then aggregated across contexts for each individual using exposure-time weights (e.g., weekly hours spent in each context).  This procedure provides a theoretically grounded way to separate lack of opportunity (low $\mathrm{EXP}$, as measured by the potential for connection in a random-mixing world) from behavioral bias ($\phi$, as measured by the shortfall from that potential). This separation is built directly into the construction of our key independent variables.

%-------------------------------------------------------------------------------
\subsection{Causal mediation for structural shocks}
%-------------------------------------------------------------------------------

\noindent
Our theory posits that major life events, or structural shocks, affect outcomes through multiple pathways—both materially and by altering an individual's epistemic framework. For example, an event like an eviction has direct financial consequences, but it may also induce stress, fatalism, or distrust, which in turn affect future behavior. This subsection specifies our strategy for empirically decomposing these mechanisms using causal mediation analysis. This allows us to quantify the proportion of a shock's total effect that is transmitted through the epistemic channel ($M_i$).

For a given structural shock $E_i(t)$ (e.g., eviction, job loss), we aim to disentangle its direct effect from its indirect effect operating through the epistemic state $M_i$. We will estimate the Average Causal Mediation Effect (ACME), Average Direct Effect (ADE), Total Effect, and Proportion Mediated using the framework developed by Imai, Keele, and Tingley. This approach relies on an assumption of sequential ignorability, and we will assess its robustness with a sensitivity analysis based on the parameter $\rho_{\nu\mu}$, which quantifies the degree of confounding needed to erase the findings. If a valid instrument for the shock $E_i$ exists, we can employ IV-mediated effects for more robust identification. This approach is supported by prior evidence, such as findings that perceived stress mediates part of the eviction–depression link \citep{Hoke2021}.  This analytical step is crucial for testing Axiom A1. It allows us to formally test the hypothesis that shocks act both materially and epistemically, and to quantify the relative importance of the frame-based pathway.

%-------------------------------------------------------------------------------
\subsection{Replication and transparency checklist (TOP/FAIR)}
%-------------------------------------------------------------------------------

\noindent
Finally, a scientific model should not only be testable but also transparent, reproducible, and open to scrutiny. This concluding section provides a checklist of commitments to open science principles, aligned with standards like the Transparency and Openness Promotion (TOP) Guidelines and FAIR Data Principles. This checklist serves as a public commitment to pre-registering our hypotheses and analysis plans, sharing our data and code, and reporting a full suite of diagnostic and robustness tests. This ensures that any empirical work conducted under this framework is maximally credible and contributes to a cumulative scientific enterprise.

\begin{itemize}
\item \textbf{Pre-registration}: All hypotheses, target estimands $(\theta_1,\theta_2,\Theta_3)$, sensitivity bounds $(\bar{\omega},\bar{\lambda},L)$, and primary analysis decisions will be pre-registered in a public repository before analysis.
\item \textbf{Data \& code}: Upon publication, we will make all necessary data and code available via a persistent digital object identifier (DOI) on a repository like OSF, Dataverse, or Zenodo. This includes scripts for (i) network calibration ($W$), (ii) counterfactual mixing ($\mathrm{EC}^{mix}$), (iii) IRT/CFA measurement models and invariance tests, (iv) IV/RDD/DiD estimation, and (v) mediation analysis.
\item \textbf{Diagnostics}: All mandatory diagnostics will be reported, including first-stage $F$-statistics, Hansen $J$-tests for overidentification, McCrary density tests for sorting, event-study pre-trends, CCT bandwidth sensitivity plots, and FDR-adjusted results for multiple hypothesis tests.
\item \textbf{Robustness}: We will report a battery of robustness checks, including the effects of winsorization or trimming for outlier values of $\mathrm{EC}$ and weights, the use of alternative functional forms for the $\Psi$ specification, and the full sensitivity analysis for the partial identification of peer effects ($\Theta_3$).
\end{itemize}

This formal model provides a rigorous blueprint for understanding and testing a complex social theory. The core argument is that inequality persists not only because people are structurally separated by class, but because deeply ingrained social norms guide who befriends whom within those structures, and because individuals' worldviews (their epistemic frameworks) are sticky—especially in adulthood. A simple policy that just "mixes" people in the same space is likely insufficient because it misses the crucial roles of friending bias and epistemic inertia. Our research program is therefore designed to test these distinct mechanisms separately: we use counterfactuals to isolate exposure from bias, we employ advanced psychometrics to measure latent frames carefully, and we leverage quasi-experimental designs that can distinguish the effects of structural opportunity from endogenous peer influence. This comprehensive approach aims to provide a more nuanced and actionable understanding of the reproduction of social inequality.

\section{Conclusion: implications of the synthetic theory}

The theory of socio-epistemic structuration offers a unified framework for understanding the complex interplay between social structures, interpretive frameworks, and individual agency in the production and reproduction of social life. By synthesizing findings from seemingly disparate fields, the theory holds that social inequality cannot be fully understood without analyzing the epistemic frameworks that legitimize, naturalize, and guide actors' responses to it. These frameworks, ranging from racial ideologies to the attribution of agency to AI, are the terrain where structure is translated into action and where action, in turn, can challenge or reinforce structure.

\begin{quote}
Critical race theories of ideology, surprisingly, replicate key tensions in early Marxist thought. Following a selective review of several Marxist theorists, this article examines critical race conceptions of racial ideology within the influential work of Charles W. Mills and Eduardo Bonilla-Silva, probing three related problem areas: an ``expressive'' theory of society, the correspondence of ideologies to racial groups, and an overreliance on standpoint implications.
\end{quote}
This quote from \citet[p.~27]{Sites2025} encapsulates a central theme of our theory: the need to move beyond ``expressive'' theories where epistemic frameworks are mere reflections of structure. Instead, we must see them as relatively autonomous and contingent constructs. Big data research reinforces this view. Economic connectedness is not an epiphenomenon but a structural variable with immense causal power over mobility. At the same time, friending bias shows that structure alone (exposure) is not destiny; micro-level choices and practices, guided by cultural frameworks, powerfully mediate structure's effect. The unpredictability of \citet{Salganik2006}'s cultural markets is a microcosm of this contingency: small initial fluctuations, amplified by social contagion, can lead to drastically different cultural outcomes.

The final implication is both methodological and political. Methodologically, we need approaches that can handle the complexity of dynamic social systems, disentangle homophily from contagion, and be humble in the face of ``big data hubris.'' Politically, the theory suggests that the most effective interventions for reducing inequality cannot focus solely on the redistribution of material resources or the reconfiguration of networks. They must actively address socio-epistemic frameworks. This means supporting institutions that reduce ``friending bias,'' creating spaces for ``reverse tutelage'' where the theories of the oppressed can gain traction, and fostering a public sociology that not only informs the public but is willing to be transformed by it. In a world where society's ability ``to make itself a god or to create gods'' is more evident than ever, as Durkheim noted of the French Revolution, the task of sociology is to understand and, when necessary, to challenge the epistemic frameworks that animate those gods.

\section*{Acknowledgments}
I used an AI tool (ChatGPT) for English translation and expository polishing of the “Formalization of the Theory” section
(adding introductory/concluding paragraphs and improving tables with \texttt{tabularx}). All analytical content, results,
and references were independently verified by the author. The AI is not an author and bears no responsibility.

\end{document}